\documentclass[aip,rsi,reprint,graphicx]{revtex4-1} 
\usepackage{graphicx}
\usepackage{amsmath}

\draft 

\begin{document}


\title{A Compact Cryogenic Thermal Source for Detector Array Characterization} 



\author{David T. Chuss}
\affiliation{Department of Physics, Villanova University, 800 E. Lancaster Ave., Villanova, PA 19085}
\email[]{david.chuss@villanova.edu}

\author{Karwan Rostem}
\author{Edward J. Wollack}
\affiliation{Code 665, NASA Goddard Space Flight Center, Greenbelt, MD 20771}
\author{Leah Berman}
\affiliation{Department of Physics, Villanova University, 800 E. Lancaster Ave., Villanova, PA 19085}
\author{Felipe Colazo}
\affiliation{Code 665, NASA Goddard Space Flight Center, Greenbelt, MD 20771}
\author{Martin DeGeorge}
\affiliation{Department of Physics, Villanova University, 800 E. Lancaster Ave., Villanova, PA 19085}
\author{Kyle Helson}
\affiliation{Code 665, NASA Goddard Space Flight Center, Greenbelt, MD 20771}
\author{Marco Sagliocca}
\affiliation{Code 665, NASA Goddard Space Flight Center, Greenbelt, MD 20771}
\altaffiliation{Department of Physics, Villanova University, 800 E. Lancaster Ave., Villanova, PA 19085}


\date{\today}

\begin{abstract}
We describe the design, fabrication, and validation of a cryogenically-compatible quasioptical thermal source designed to be used for characterization of detector arrays.  The source is constructed using a graphite-loaded epoxy mixture that is molded into a tiled pyramidal structure. The mold is fabricated using a hardened steel template produced via a wire EDM process.  The absorptive mixture is bonded to a copper backplate enabling thermalization of the entire structure. The source reflectance is measured from 30-300 GHz and compared to models. \end{abstract}


\maketitle 

\section{INTRODUCTION}
\label{sec:intro}  


The effort to develop large focal planes operating in the far-infrared through millimeter part of the spectrum requires the capability to test and calibrate the optical response of the sensors. Absolute detector efficiency must be tied to a known calibration source.  Such calibration can be done outside of the instrument; however, in this configuration, the measurement determines the total efficiency of the instrument and not just the detector.  In validating a sensor design, isolation of the detector response from that of the instrument can provide a valuable diagnostic tool. 

Epoxies loaded with conductive material \cite{Wollack08, Persky99} have been used as thermal calibrators both because of their electromagnetic properties and their compatibility with casting processes. This enables the geometry to be used to enhance the total absorptance and to constrain the thermal profile. For individual channels, single-mode waveguide loads have been demonstrated to provide precise calibration standards\cite{Wollack07,Rostem13}.  However, this can only be realized on the basis of a single (dual polarization) waveguide input per calibrator, so characterizing a large array of detectors in this manner can be impractical.  Multi-mode high precision broadband devices have been employed in the COBE/FIRAS\cite{Mather99} mission and for the ARCADE balloon payload\cite{Fixsen06} as calibration standards for absolute measurement of the cosmic microwave background. Similar principles have been applied in the context of millimeter and submilimeter wave remote sensing\cite{Carli74,Betts85,Wollack16}. 

In this application, we pursue a monolithic design suitable for validation of large arrays of cryogenic sensors. 
We present the design, fabrication, and validation of a free-space cryogenic calibrator that can be employed to measure a large number of detectors simultaneously. The calibrator is a cast loaded epoxy structure consisting of a Cartesian tiling of nominally identical pyramids, each having full angle of 20$^\circ$. A copper backplate provides a thermalization layer as well as attachments for thermometry, heaters, and mechanical support. The cone's impedance profile\cite{Colin} is determined both by the materials used and the detailed geometry of the taper.  The impedance profile in turn defines the reflectance. The symmetry of the absorber and tiling also has an influence on the polarization response\cite{Mackay89}.  In this work, a calibrator approximating a four-fold symmetric pyramidal tiling is employed to mitigate on-axis cross-polarization.


\section{MODELING}
\begin{figure} [ht]
   \begin{center}
   \begin{tabular}{c} 
   \includegraphics[width=3in,angle=0]{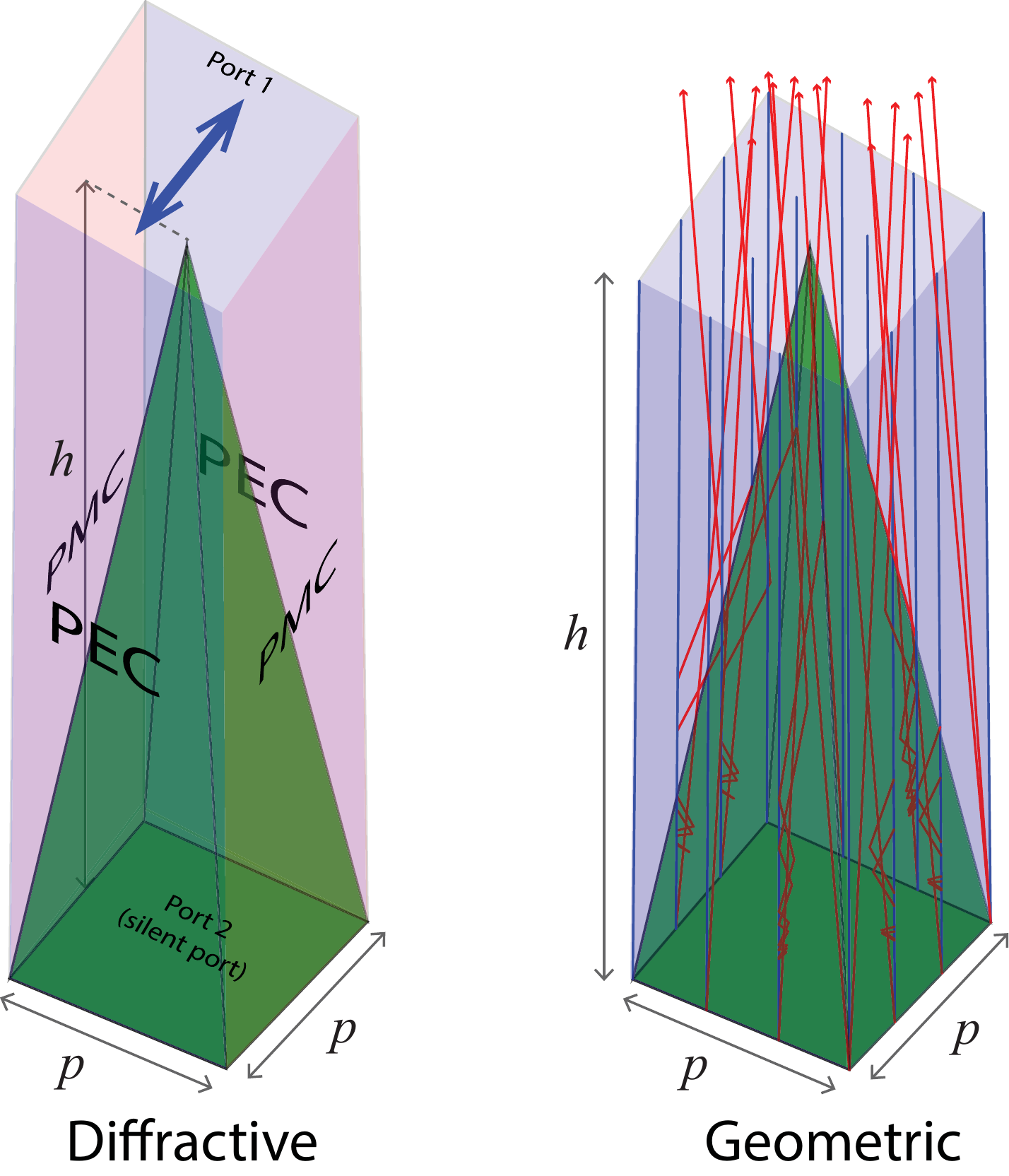}
   \end{tabular}
   \end{center}
   \caption[template] 
   { \label{fig:gmod1} 
   The geometry of the unit cell for the absorber is shown for both the diffractive (left) and geometric (right) models.  For the diffractive model, port 1 is set to transmit and receive. Port 2 is designated to only receive (silent port). The polarization direction is indicated by the blue line at the top of the structure. For the geometric model, the blue (parallel) rays are the input rays. The exit rays are shown as red arrows. The polarization is tracked through the system. In each case, the boundary conditions are periodic.  For the diffractive model, this is accomplished with perfect electric conductor (PEC) and perfect magnetic conductor (PMC) boundaries as shown. For the geometric case, rays intercepting a boundary wall are mapped to the opposite boundary (see Appendix for details).}
   \end{figure}

The modeling for the devices is performed in two distinct regimes. In the diffractive regime, where few modes exist within the structure, a finite element analysis (FEA) electromagnetic model is used. As the simulation frequency is increased and the total number of modes rapidly expands, computation limits are reached with this approach. In the geometric optics limit, a computationally efficient model that take the dominate vector nature of the polarization into account is employed. Figure~\ref{fig:gmod1} provides an illustration of the unit cell for the absorber geometry and considerations for the models used in both the diffractive and geometric limits.

\subsection{The Diffractive Limit}
When the wavelength similar to or greater than the pitch of the absorber tiling, the electrical size of the model enables the active modes in the problem to be modeled using finite element analysis. Here the COMSOL Multiphysics software package is employed for this purpose in the simulations performed. This is a full-wave electromagnetic finite element analysis in which a single pyramid as defined by its apex angle is used as the unit cell. Periodic boundary conditions are implemented. This modeling technique is used to explore the parameter space and to converge on an acceptable manufacturable geometry.

The power reflectance, $R$, is modeled parametrically as a function of the ratio of the pyramid height, $h$, to the pitch of the tiling, $p$. In cryogenic applications, the desired to minimize the magnitude of $R$ and maximize total absorptance must be considered in the context of the thermal performance. For example, the entire volume of the absorber material of the calibrator will determine the heat capacity. A larger taper ratio $h/p$ decreases the reflectance, but also increases the temperature gradient across the taper, leading to a bias in the calibrator temperature due to the finite thermal conductivity of the absorber material. In general, the thermal conductivity and dielectric loss of the absorber material are positively correlated. In addition, longer tapers are more challenging to manufacture.

\begin{figure} [ht]
   \begin{center}
   \begin{tabular}{c} 
   \includegraphics[width=3.4in,angle=0]{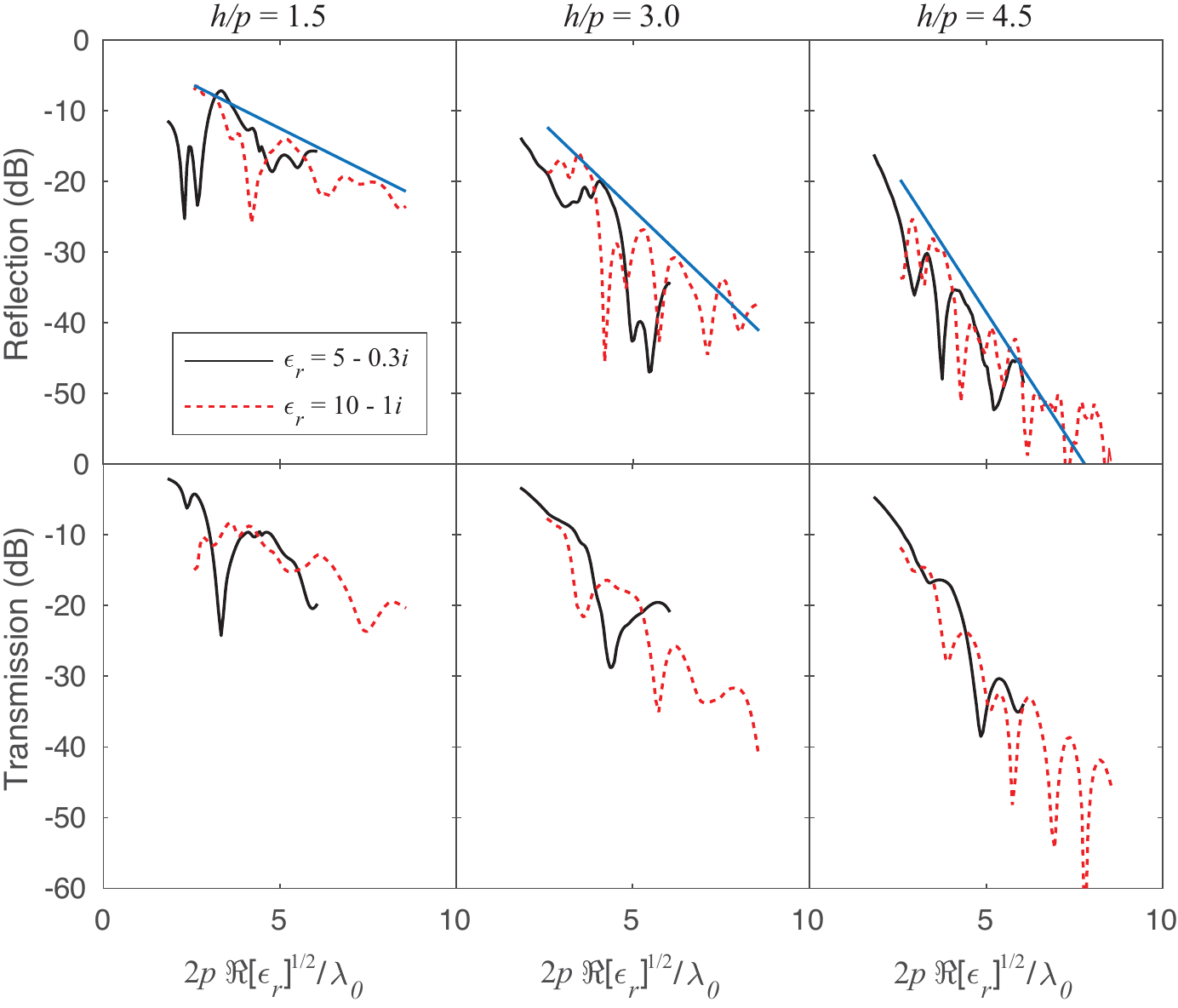}
   \end{tabular}
   \end{center}
   \caption[template] 
   { \label{fig:dmod} 
   Reflection and transmission are shown as a function of scaled frequency for three values of the ratio of the height ($h$) to the pitch ($p$) of the pyramidal tiling.  In each case, models are shown for $\epsilon_r=5-0.3i$ and $\epsilon_r=10-1i$. The loss tangent in each case is constant. The envelope for the diffractive limit is shown as a blue line in each of the top three plots.
  }
   \end{figure}

At normal incidence, an infinite tiling of pyramidal absorbers is defined by perfect electric conductor (PEC) and perfect magnetic conductor (PMC) boundary conditions. The simulations are performed for a single polarization as defined the unit cell boundaries. For simulations with off-axis plane wave illumination on the absorber, periodic boundary conditions are used. In Fig.~\ref{fig:dmod}, we use a perfectly matched silent port at the base of the pyramid, such that the structure's reflectance, R, and the transmittance, T, are monitored and the absorptance, $A = 1-T-R$, can derived from the simulation. In simulating the desired absorber configuration with a metal backing, a PEC boundary condition is employed at the base of the pyramid. These two configurations can provide insight into the required absorber volume and the role of coherent reflections from and within the structure. The separation between the model geometry and the waveport is  greater than 3 times the maximum freespace wavelength to avoid spurious interactions over the simulation dynamic range of interest.  In all simulations, the maximum mesh element size in our FEM models does not exceed $\lambda_g/(5\,\sqrt{\Re[\epsilon_r]})$. The convergence criteria is $<-40$ dB in power. Here, $\lambda_g$ is the guide wavelength in the absorber structure's dielectric media.

It is instructive to consider the antenna theory analogue of a tapered absorber structure. It can be shown that there is a one-to-one mathematical correspondence between the angular response of an apodized illumination antenna function and the reflection of an adiabatic absorber structure~\cite{Matthaei}. This motivates use of the following parameterizing~\cite{Jasik} for the reflectance, $R = 1/[1+(\lambda_g/\lambda_{c})^d]$, where the exponent $d$ is a function of the absorber impedance taper, $\lambda_g = \lambda_o/\sqrt{\Re[\epsilon_r]}$ is the wavelength inside the absorber material, and $\lambda_c = 2p$ is the cutoff wavelength of a tapered load in free space. These scaling relations are used to present the data shown in Figure~\ref{fig:dmod}. As expected, the envelope response for a given absorber taper is not dependent on the dielectric function of the media in this representation. This representation enables the trade between the reflectance and physically required volume for a given taper aspect ratio to be seen.

\subsection{The Geometric Limit} 
When the wavelength is small compared to the scale of the pyramid pitch, many modes are present, and a geometric model can be used to approximate the reflectance 
of the calibrator.  In this model, the light is geometrically traced through a unit cell of absorber. At each interaction between the ray and the calibrator structure, the absorption is determined by the Fresnel equations and is therefore parameterized by the incidence angle, polarization state and by the permittivity of the material.  Each reflection is treated as incoherent with respect to the other reflections to reduce the dimensionality of the problem and to enable a tractable numerical formulation in the large mode limit. A ray that is transmitted into the material is assumed to be completely dissipated. 

The geometry of the model is shown in Figure~\ref{fig:gmod1}.  A unit cell is composed of a pyramid with height $h$ and a square base with a side corresponding to the pitch, $p$. Four virtual walls are placed at the symmetry boundaries of the unit cell.  As a ray propagates into the structure, its position and direction are tracked. After each interaction between a ray and one of the eight planes (4 planes representing the cone and 4 planes representing the boundary conditions), the distance between the current position of the ray and each of other planes is calculated.  The ray is then propagated to the surface that is closest to the current position. If this surface is a boundary wall, the position of the ray is translated to the opposite wall, and the angle of the propagation is unchanged. This can be thought of as simply replacing a ray that is exiting the unit cell with an equivalent ray entering the unit cell from its opposite neighbor. When the nearest surface is one of the pyramid sides, the ray is first propagated to the cone surface. The direction of the ray is then adjusted according to the law of reflection at that surface. The polarization is rotated into a basis whose vectors are parallel and perpendicular to the local plane of incidence, and the Fresnel coefficients are used to 
determine the attenuation of each of the polarizations. This process continues until the ray crosses the perpendicular plane at the cone's tip. 

For each incidence angle, the process is repeated with an array of rays at starting points equally distributed across the unit cell. The mathematical formulation used in this model is described in Appendix A.
\begin{figure} [ht]
   \begin{center}
   \begin{tabular}{c} 
   \includegraphics[width=3.3in,angle=0]{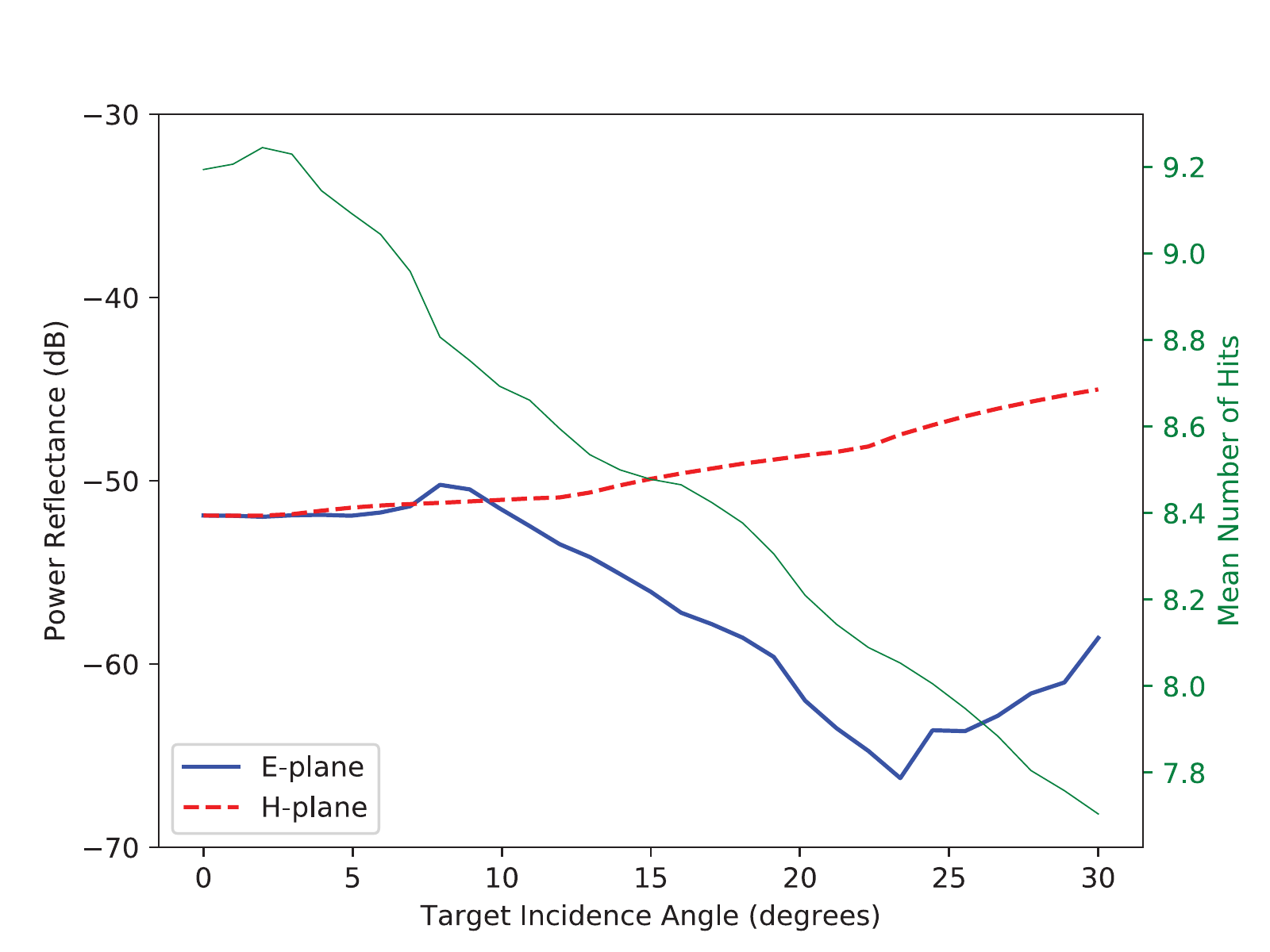}\\
    \includegraphics[width=3.3in,angle=0]{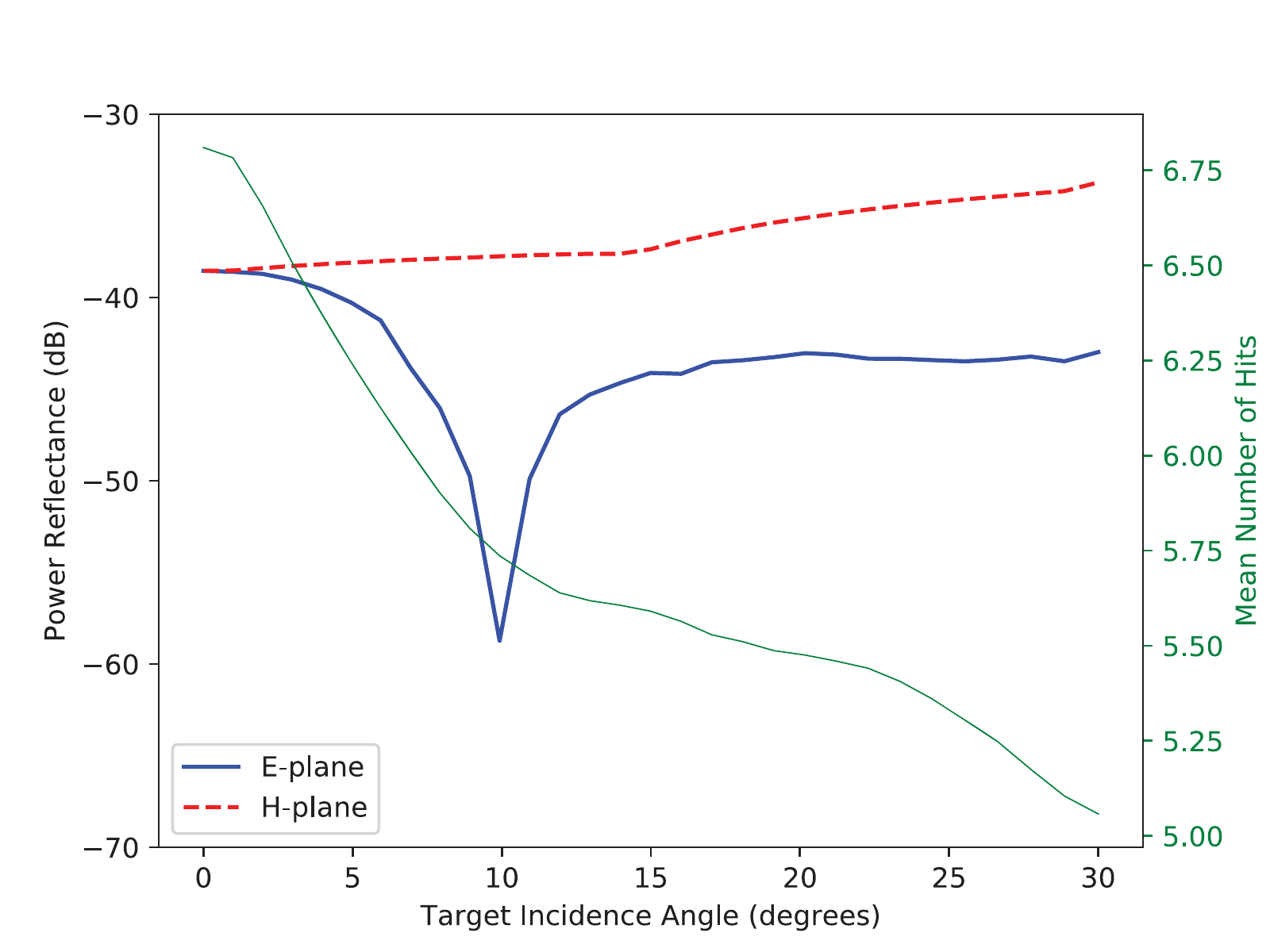}
   \end{tabular}
   \end{center}
   \caption[template] 
   { \label{fig:gmod} 
   The geometric limit model reflectance for the polarization parallel to the E- and H-plane is shown. In each simulation, 10,000 equally-spaced rays were traced through the system at 1 degree angle increments. The mean number of interactions with the wall are also shown as a function of target incidence angle. For the top model,  $\epsilon_r=5.2-0.3i.,$ $h=11.7$ mm, and $p=4.0$ mm. For the bottom model the following parameters are adopted: $\epsilon_r=4.15-0.66i,$ $h=7.4$ mm, and $p=4.0$ mm.} 
   \end{figure}    
In Figure~\ref{fig:gmod}, the reflectance is calculated in the geometric limit as a function of incidence angle for two different geometries. Both the E-plane and H-plane responses are shown.

\begin{figure} [ht]
   \begin{center}
   \begin{tabular}{c} 
   \includegraphics[width=3.5in,angle=0]{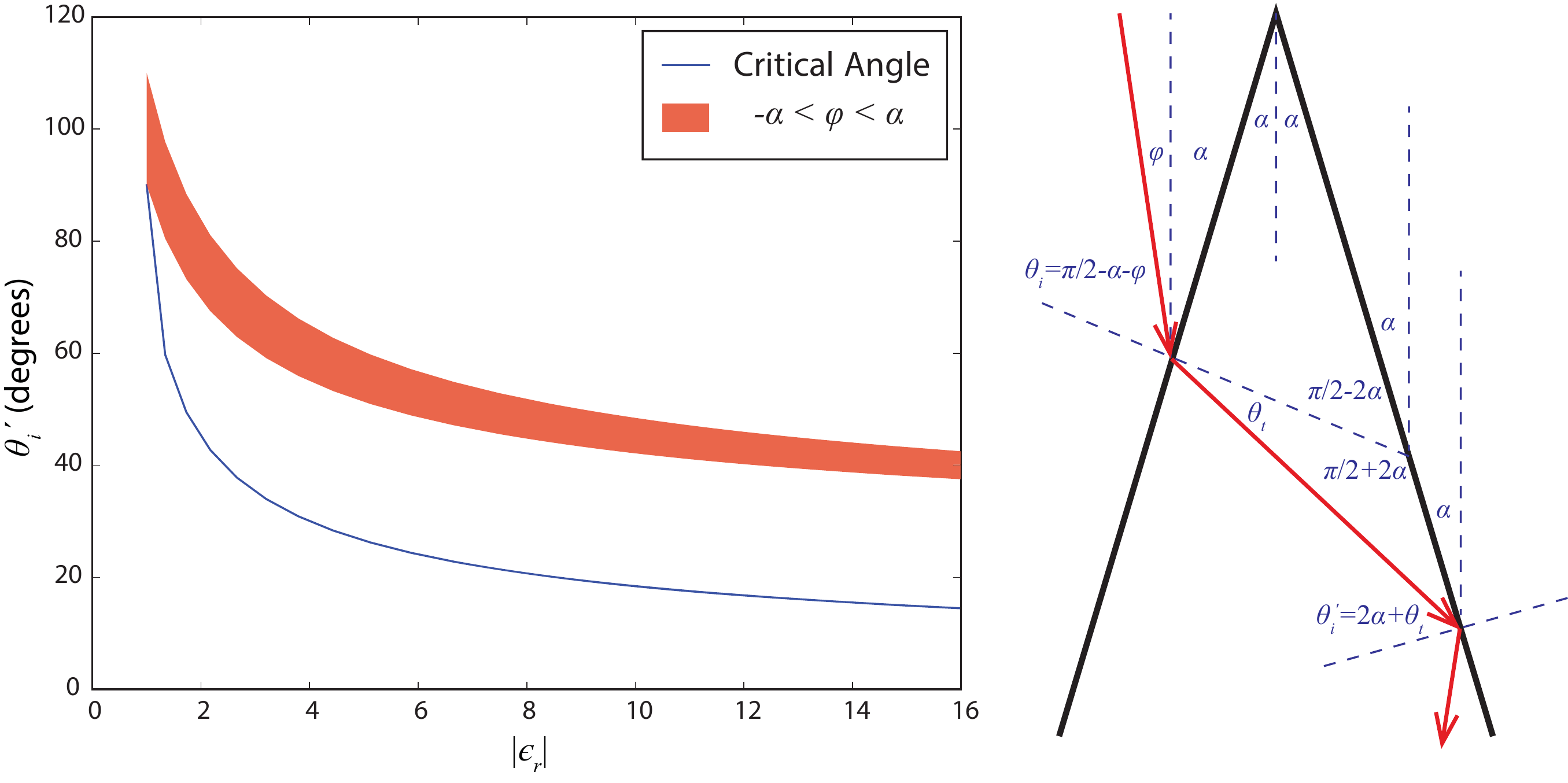}
   \end{tabular}
   \end{center}
   \caption[template] 
   { \label{fig:cang} (Left) The incidence angle for a ray exiting the cone is shown as a function of $|\epsilon_r|$ over a range of incidence angles up to half the pyramid angle. For all of these rays, the incidence angle at the second surface of the cone exceeds the critical angle. For this plot, $\alpha=10^\circ$. (Right) The geometry of the setup is shown.
   } 
   \end{figure}

The geometric model relies on the assumption that any ray that gets into the loaded dielectric is absorbed completely. One potential violation of this assumption occurs near the tip of the pyramid where the path through the absorber is short. In this case, we show (see Fig.~\ref{fig:cang}) that for target incidence angles having magnitudes up to half the pyramid angle (assuming a pyramid half-angle of 10$^\circ$), the incidence angle at the exit surface exceeds the critical angle for a large range of permittivity values. Thus, rays not terminated within their path through the absorber are totally internally reflected for relevant directions.

\section{FABRICATION}

\subsection{Target Parameters}
From the simulations in Figure~\ref{fig:dmod}, we selected a pyramidal absorber with $p\simeq4$ mm, $h=11.7$ mm ($h/p \sim 3$). This choice is informed by the target reflectance of -20dB at 75-90 GHz, manufacturability (see Sec.~\ref{sec:mold}), and the dielectric function of the absorber materials available that could be cast to form a pyramidal absorber array. For the latter, the material loss and ease of mixing and molding to shape are major considerations. We use the dielectric mixture described in Table~\ref{tab:377}, which utilizes a Epotek 377 epoxy binder with SiO$_x$ to improve the match to the thermal expansion of the copper backplate, and graphite to provide conductive loading. This mixture was selected among several candidates because its low viscosity minimizes the formation of voids in the mixture. This leads to an improved surface finish and specular reflection from the final product. The volume and mass fractions used are given in Table~\ref{tab:377}. This material approximates the properties of the lower $\epsilon_r$ material in Figure~\ref{fig:dmod}.

\begin{table}[htbp]
   \centering
   \begin{tabular}{lcc} 
   	\hline
	Material & Volume Fraction (\%) & Mass Fraction (\%) \\\hline
	Epotek 377 & 65 & 50.7\\
	SiO$_x$ & 30 & 42.8\\
	Graphite & 5 & 6.5\\\hline
   \end{tabular}
   \caption{Components of the Thermal Source by Volume and Mass Fractions}
   \label{tab:377}
\end{table}


To ensure that the device adequately approximates an isothermal source, we utilize FEA simulations assuming that the highly conductive copper base is isothermal. The conductivity of the epoxy mixture is anticipated to be similar to that of Stycast 1266 \cite{Olson}. The results of our model indicate that for a backplate temperature of 15 K, the temperature gradient along the longest dimension of the pyramid is $\ll1$ mK, which has negligible impact for the end application. 

\subsection{Creating the Mold}
\label{sec:mold}

\begin{figure} [ht]
   \begin{center}
   \includegraphics[width=3.3in,angle=0]{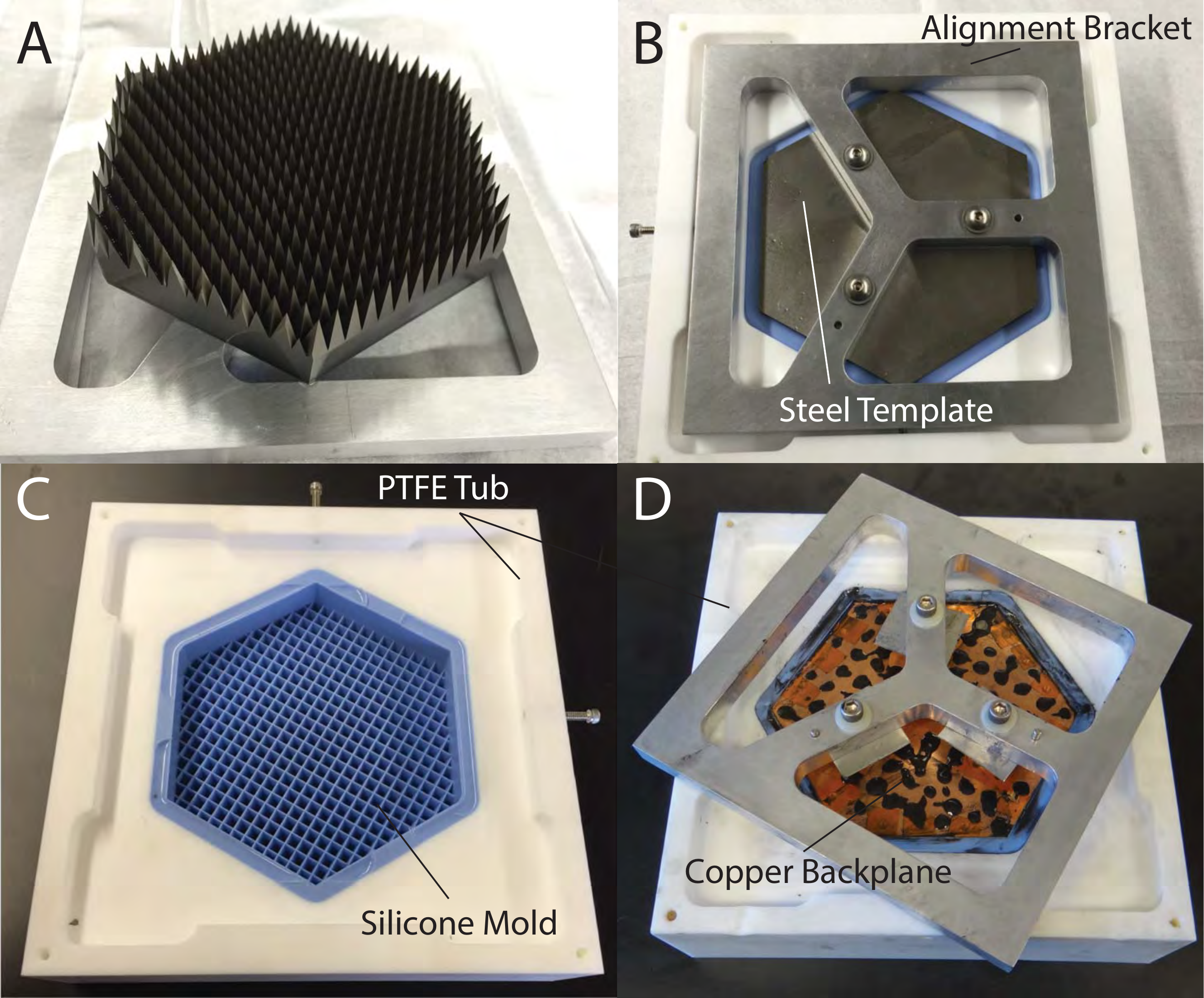}
   \end{center}
   \caption[template] 
   { \label{fig:template} 
   A. The template is shown attached to the alignment bracket. The pyramids are created via wire EDM using hardened steel. B. The template is suspended in the PTFE tub via an attached alignment bracket and secured against reference edges using a pair of screws.  C. Once cured, the template is removed from the mold. D. he copper backplate is suspended in the epoxy using the same alignment bracket as for molding.  Small through-holes in the backplate allow epoxy to flow through. This serves two purposes. First, the holes allow any trapped air to escape. Second, the holes promote mechanical adhesion between the epoxy and the backplate.}
   \end{figure}

The mold positive, or template, is constructed via wire Electrical Discharge Machining (EDM) using a 250 $\mu$m diameter wire.  Hardened steel is used and an array of pyramids is cut into a hexagonal block.  Screw holes and pins on the back of the template allow alignment and mechanical connection to a mounting bracket.  This technique has been found to produce tips that are sharper than 10 $\mu$m. Figure~\ref{fig:template} shows the wire EDM machined template. 

The template is used to make a mold using silicone.  The silicone mixture is poured into a PTFE tub. The template is suspended in the bath using a bracket, which is used to center the template using reference surfaces in the teflon tub and screws to secure the bracket in place.

After a 24-hour room temperature cure, the Silicone mold is removed from the template and baked at $\sim 100^\circ$ C for one hour to finish the curing process. The molding process is shown in Figure~\ref{fig:template}.
 
\subsection{Casting the Absorber}
The absorber is cast in the same PTFE tub as the silicone mold. A commercially-available silicone mold release is used to coat the silicone mold to enable the calibrator to be removed after curing. The alignment bracket is used to suspend the copper backplate at the surface of the loaded epoxy. The bracket is lowered until the epoxy fills a grid of through-holes in the backplate.  These perforations allow the release of trapped air and ensure good mechanical adhesion between the copper and the absorbing material. 

\begin{figure} [ht]
   \begin{center}
   \begin{tabular}{cc} 
   \includegraphics[width=1.65in]{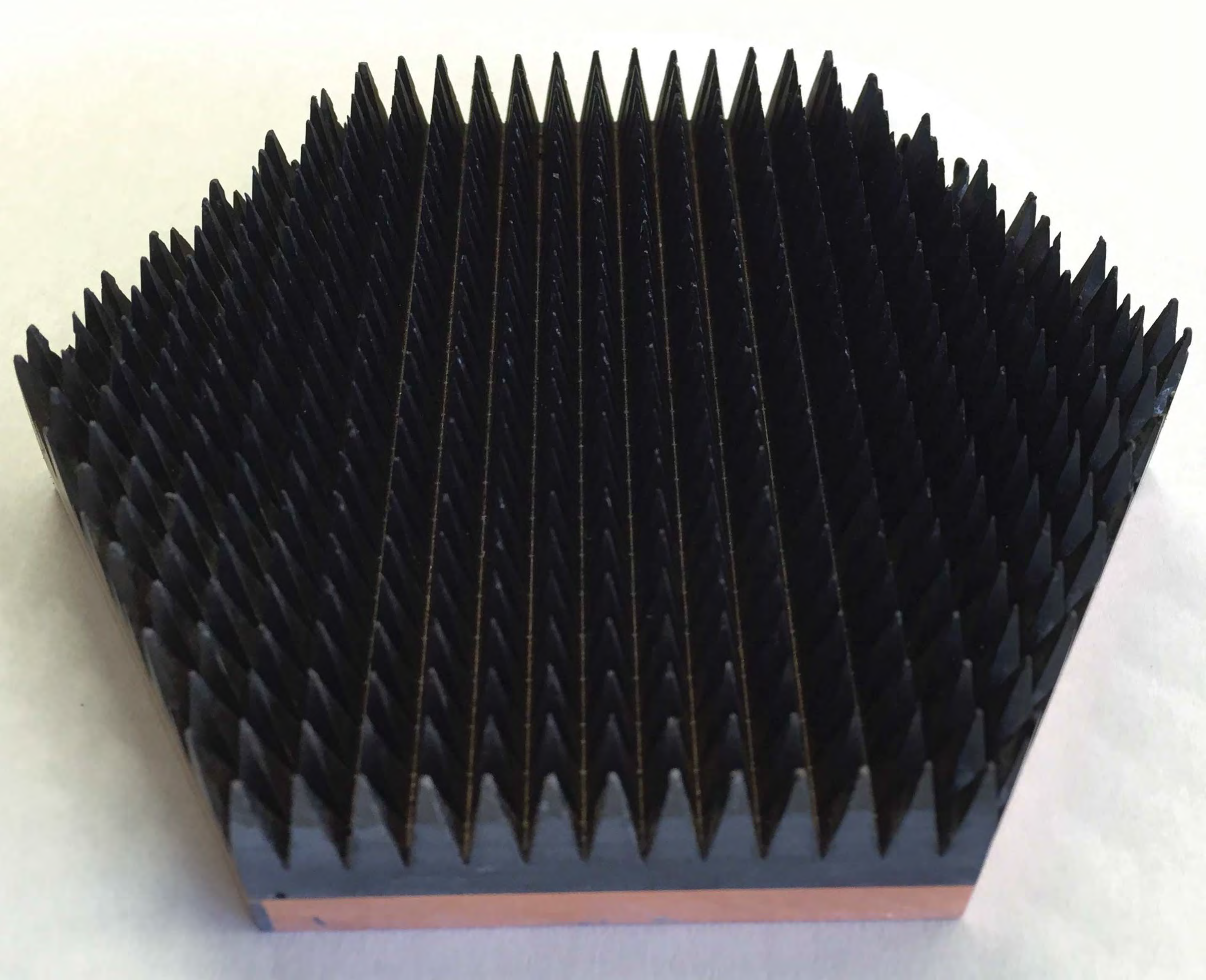} &
   \includegraphics[width=1.65in]{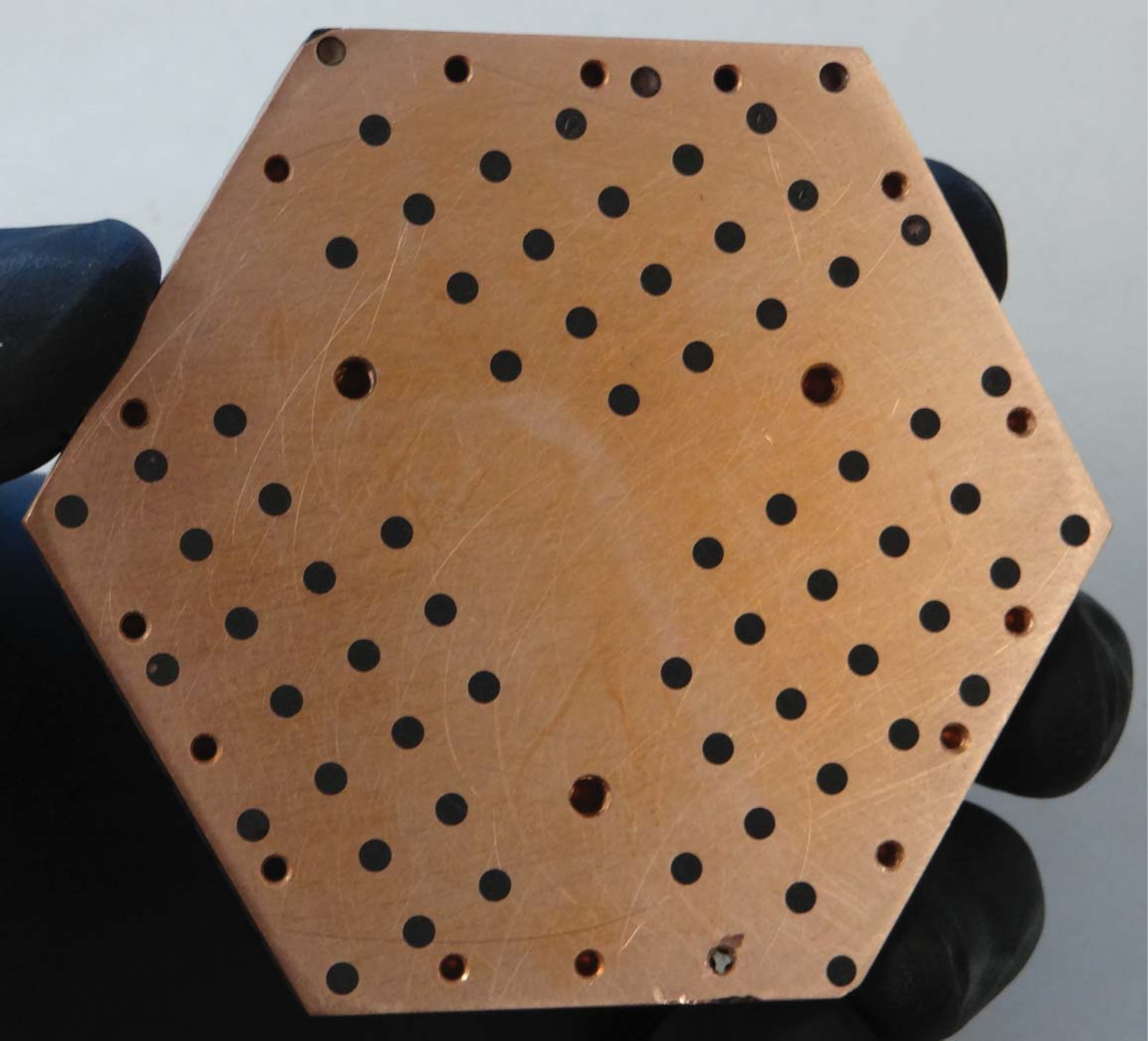}\\
   \end{tabular}
   \end{center}
   \caption[casting2] 
   { \label{fig:casting2} 
   (Left) A prototype device is shown. (Right) The back of the calibrator is lapped flat to remove excess epoxy and to enable reliable mounting in a cryogenic environment.
}
   \end{figure}

Once the epoxy is cured, the calibrator is removed from the mold. Excess epoxy around the edges of the hexagonal structure is removed by sanding. The backplate is lapped flat to ensure good thermal conduction when installed into a cryostat.  This process also removes excess epoxy at the hole locations. A prototype device is shown in Figure~\ref{fig:casting2}. The realized sizes of the tips are typically $\gtrsim$125 $\mu$m. The valleys between the pyramids are approximately cylindrical with characteristic radii of 165 $\mu$m.

\section{VALIDATION}

The complex relative dielectric function of the loaded epoxy was characterized using rectangular waveguide (WR28.0; broadwall 0.280'' and guide height 0.140'') section with a pair of filled guide sections. Each waveguide shim forms a Fabry-Perot resonator and the dielectric function is extracted from the scattering parameters measured from 22-40\,GHz with a PNA-X Vector Network Analyzer sampled at 801 points following the modeling approach previously described\cite{Wollack08}. These shims were produced as witness samples during the fabrication of the target. 
For comparison, the commercially-available Thomas Keating RAM tile is also considered. This device is an injection-molded polypropylene-based structure designed for low reflection (http://www.terahertz.co.uk).  Waveguide samples were also produced using pieces of this material cut from the edge of a tile.  Results of the dielectric characterization along with the geometry of each device are given in Table~\ref{tab:devices}. 
\begin{table}[htbp]
   \centering
   \begin{tabular}{lcccc} 
   	\hline
	Device & $\epsilon_r$ & height & pitch &$h/p$  \\
			& (-) & ($h$, mm) &  ($p$, mm)\\ \hline
	Thomas Keating RAM  & $4.15-0.66i$ & 7.4 & 4.0 & 2\\
	Thermal Source & $5.2 - 0.2i$ & 11.7 & 4.0 & 3\\\hline
   \end{tabular}
   \caption{Comparison of the Thermal Source and the Large TK RAM}
   \label{tab:devices}
\end{table}

\begin{figure} [ht]
   \begin{center}
   \includegraphics[width=3.25in]{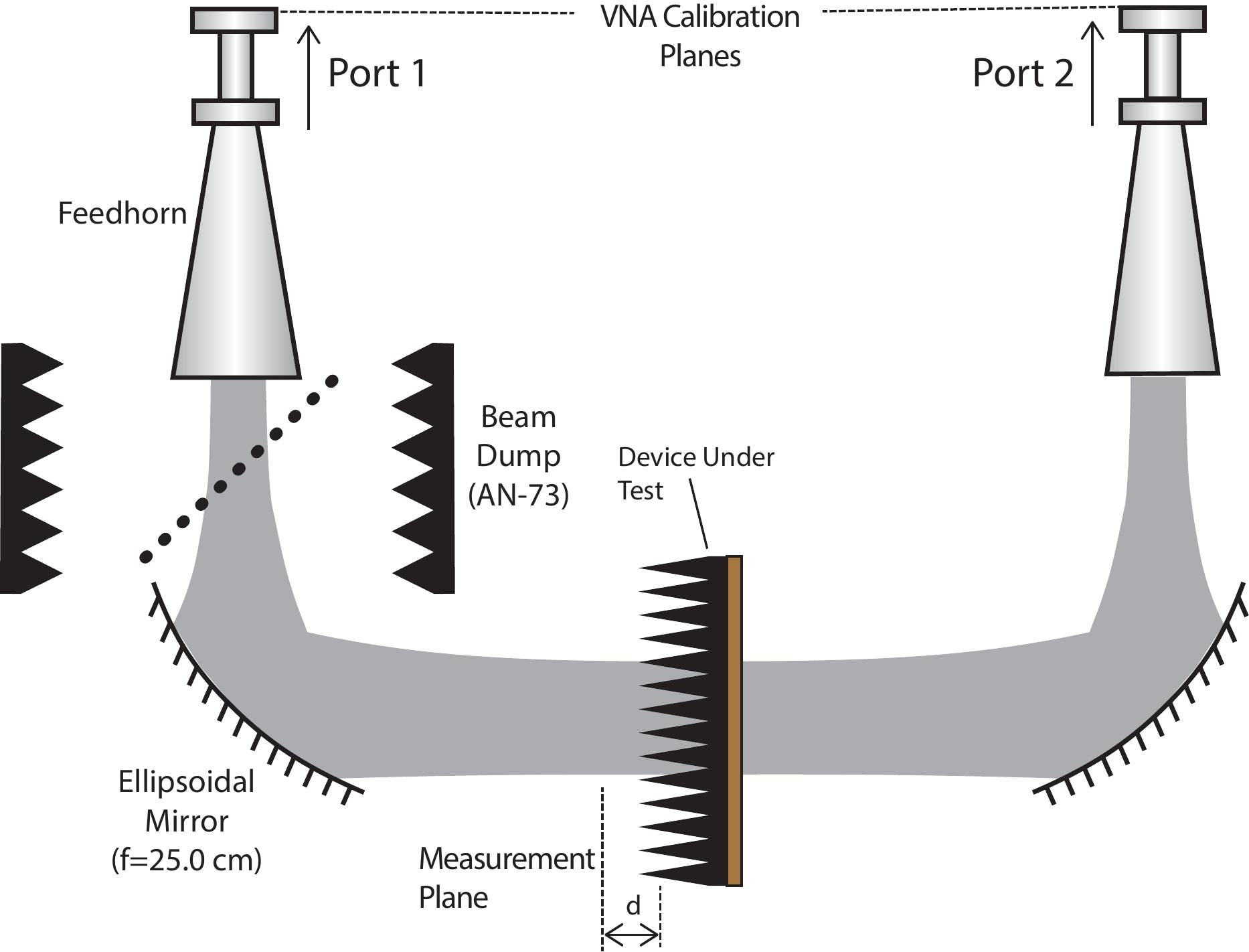}\\
   \end{center}
   \caption[setup] 
   { \label{fig:setup} 
	The setup for the measurement is shown. 
	}
   \end{figure}

\begin{figure} [ht]
   \begin{center}
   \includegraphics[width=3.3in]{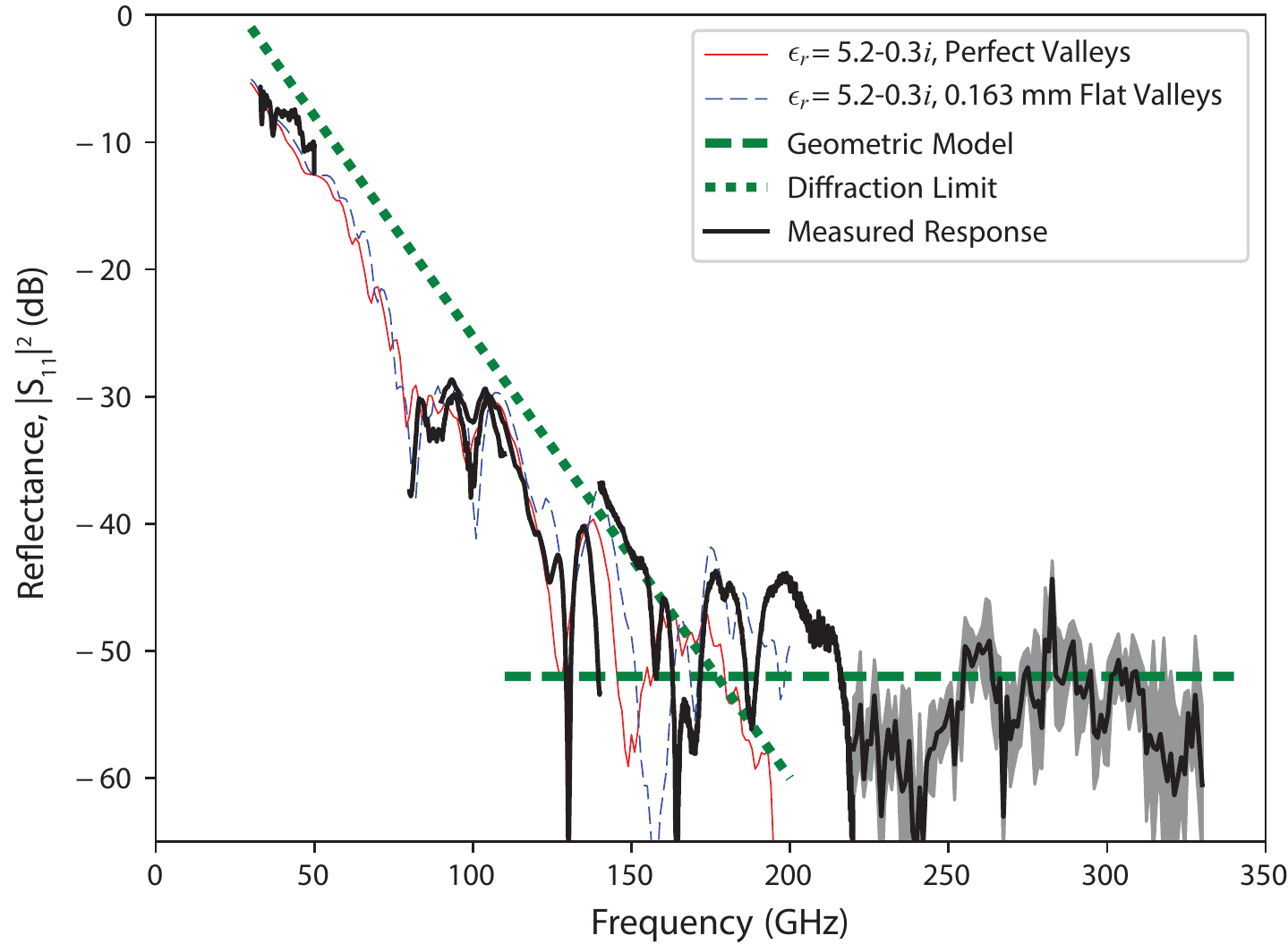}\\
   \includegraphics[width=3.3in]{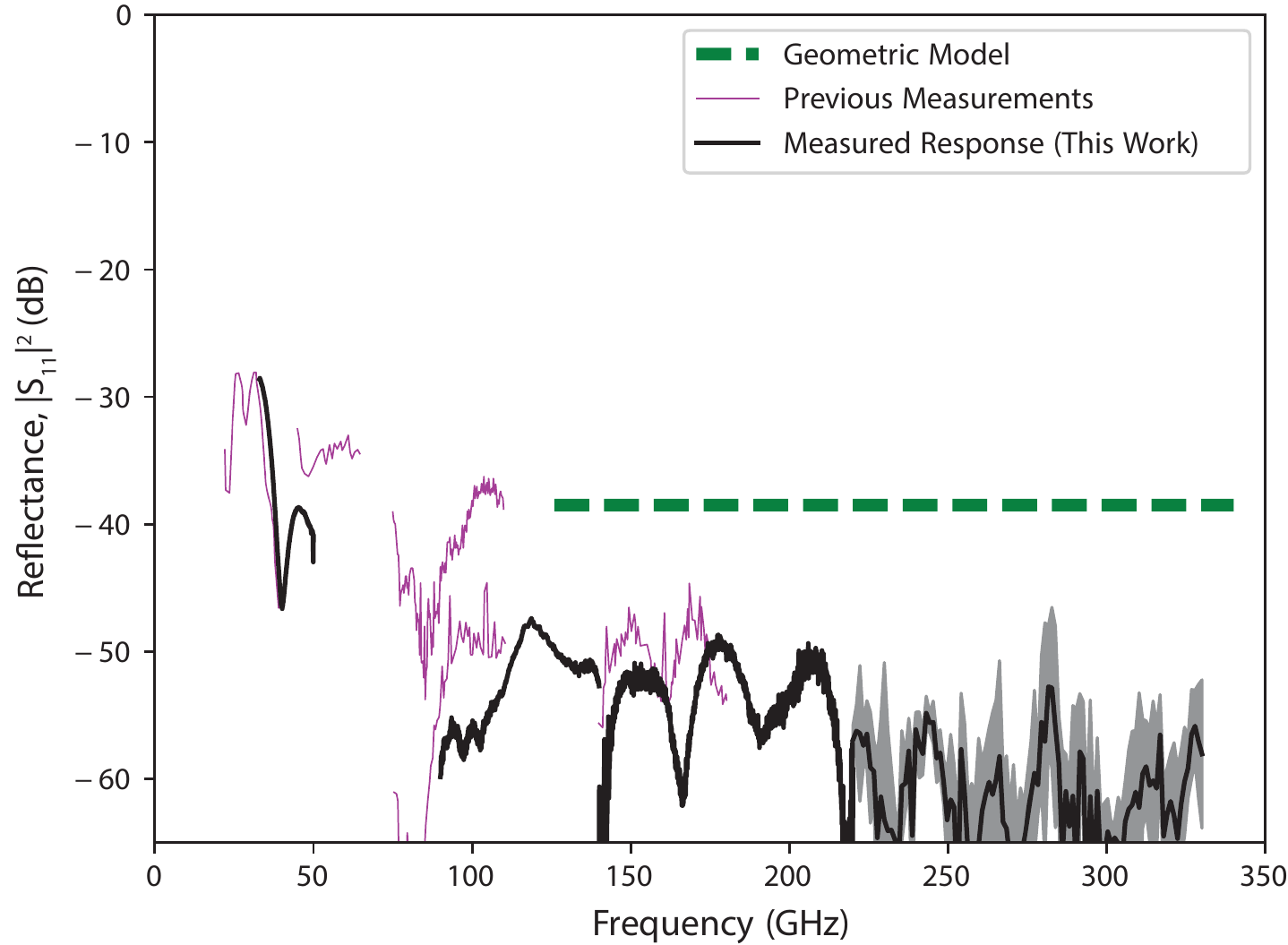}\\
   \end{center}
   \caption[results] 
   { \label{fig:results} 
(Top) Results of the prototype are shown for reflection measurements made in five separate VNA bands. Two models in the diffractive limit are indicated by ``Perfect Valleys'' and ``Flat Valleys.'' The first assumes that the valleys between the pyramid structures are perfectly sharp. The second assumes that the valleys are flat channels that are 163 $\mu$m wide. The geometric optics limit for normal target incidence as calculated and indicated (long dashed lines). (Bottom) Similar measurements for a Thomas Keating (TK)  Large RAM tile are plotted with measurement from the TK website (http://www.terahertz.co.uk) and the geometric model results for normal incidence. In both plots, the gray bands indicate the 1-$\sigma$ bounds obtained from the variance of 4 measurements in the highest waveband presented.
}
	\end{figure}

A prototype device was measured using a vector network analyzer (VNA) coupled to the free space quasioptical setup shown in Figure~\ref{fig:setup}. A multi-tier calibration is employed in the setup at each frequency. This consists of an initial waveguide calibration of the VNA using a 2-port through-reflect-line (TRL) calibration technique at the planes shown in this figure.  Next, the feedhorns are connected, and the sample is mounted on the translational sample stage. Horizontal polarization from a single VNA port (port 1) is normally incident on the calibrator.  Measurements are made at several positions, $d$, of the calibrator. Combining these measurements enables the instrument response to be separated from the calibrator reflection \cite{Eimer11}. Measurements were done in 5 separate waveguide bands to span a spectral range from 30 to 330 GHz. Results are shown in Figure~\ref{fig:results}. This figure also shows measurement data from a large TK RAM tile sample.  Data for the latter device are underestimates of the specular reflectance of a perfectly flat sample due to the observed camber of the tile. This feature defocuses the power incident from port 1 and is not an issue for the thermal source sample.

The reference level is determined by replacing the target with a flat aluminum plate and verifying. In addition, the edge illumination of the device and port-to-port isolation was also measured. This was done by terminating the boundary of the target beyond its edge with AN73 absorber and comparing to the unterminated response. The termination was found not to change the response of the calibrator, thus validating the measurement. The prototype tile has been thermally cycled to below 4 K twice and survived.  

\section{SUMMARY}
We have produced a prototype cryogenic multi-mode calibrator that will be used in characterization of arrays of millimeter-wave detectors. The devices is fabricated using a loaded epoxy that is cast into an array of pyramids onto a copper backplate for thermalization. A target reflectance of $<$0.01 has been realized a spectral range from 75 to 330 GHz, which is adequate for the desired application. The reflectance is controlled by the impedance taper in the single mode limit~\cite{Southwell91} and the geometry and surface details in the geometric limit~\cite{Wollack14, Schroeder16}. The analysis and design synthesis approach employed here allows the response to be tailored to reduce the reflectance in both the diffractive and geometric limits and has been experimentally validated with a simple linear geometric taper geometry. More complex structures are amendable to this treatment in the limit the surface roughness is small compared to incident wavelength.

\begin{acknowledgments}
The authors would like to thank Riley McCarten for help in fabricating absorbers and Berhanu Bulcha for metrology support.
We would also like to thank Paul Mirel for guidance on molding and casting. This work was funded by a NASA ROSES
Strategic Astrophysics Technology (SAT) award and by NASA Grant NNX16AC33G.
\end{acknowledgments}

\section*{Appendix: Geometric Model Details}
\begin{figure} [ht]
   \begin{center}
   \begin{tabular}{c} 
   \includegraphics[width=3.2in,angle=0]{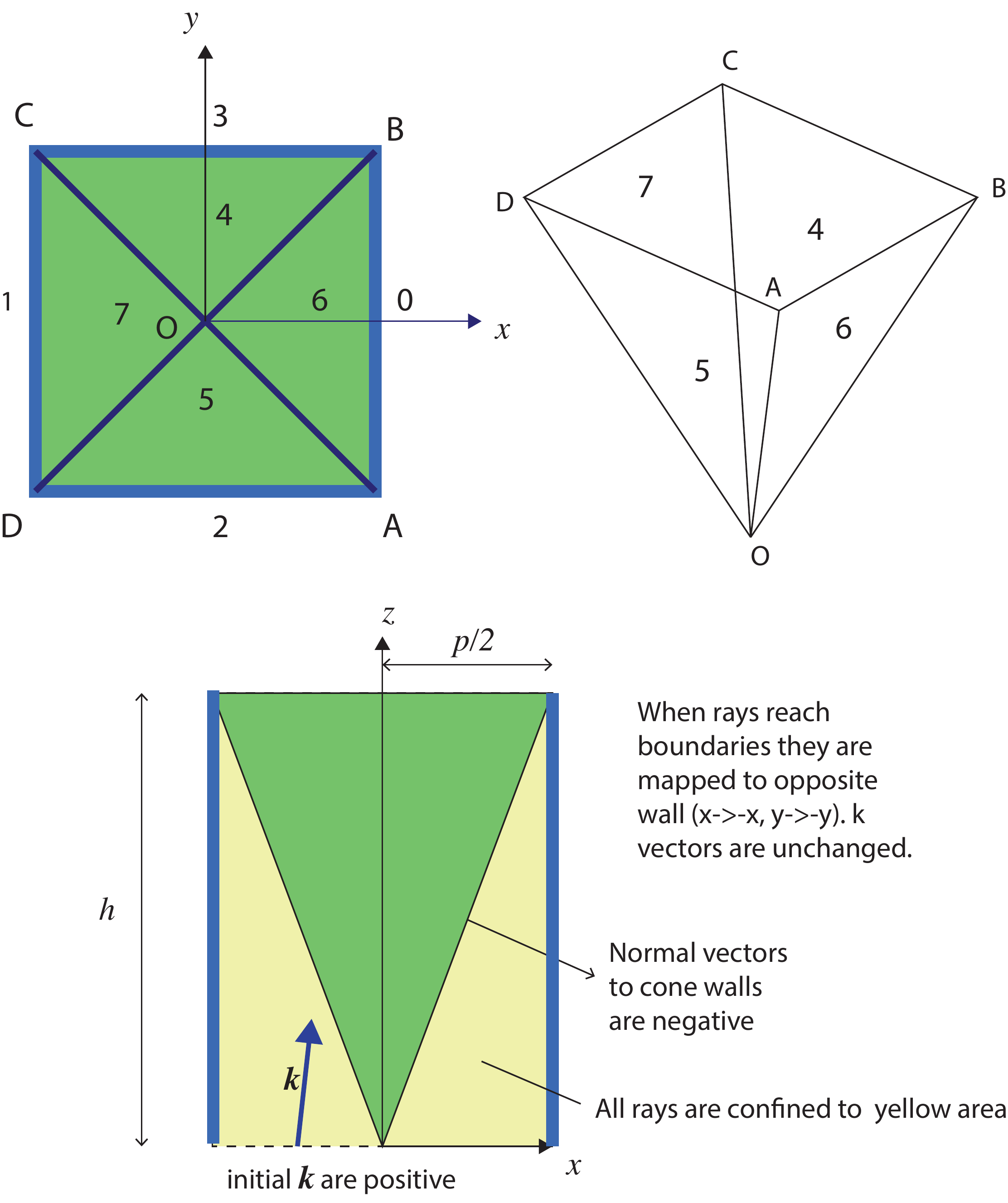}
   \end{tabular}
   \end{center}
   \caption[template] 
   { \label{fig:reff} 
    }
   \end{figure}

The coordinate system adopted for this problem is illustrated in Figure~\ref{fig:reff}.  The origin is chosen to be the tip of the pyramid, and initial rays are chosen to start in the $x$-$y$ plane and have a positive direction.  The rays will interact with one of the eight planes in the problem. The first four comprise the boundary surfaces on the planes of symmetry, and the second four make up the pyramidal absorbing structure. To define each plane, we require a point and a normal vector.  In accordance with the labeling in Figure~\ref{fig:reff}, the points for each of the 8 planes are
\begin{equation}
 \begin{array}{ll}
{\bf r}_0=\frac{p}{2}\bf{\hat{x}} & {\bf r}_4=\frac{p}{4}{\bf \hat y}+\frac{h}{2}{\bf \hat z}\\
{\bf r}_1=-\frac{p}{2}\bf{\hat{x}} & {\bf r}_5=-\frac{p}{4}{\bf\hat y}+\frac{h}{2}{\bf\hat z}\\
{\bf r}_2=-\frac{p}{2}\bf{\hat{y}} &{\bf r}_6=\frac{p}{4}{\bf\hat x}+\frac{h}{2}{\bf\hat z}\\
{\bf r}_3=\frac{p}{2}\bf{\hat{y}}& {\bf r}_7=-\frac{p}{4}{\bf\hat x}+\frac{h}{2}{\bf\hat z}.
\end{array}
\end{equation}
The normal unit vectors for each of these surfaces are
\begin{equation}
\begin{array}{ll}
{\bf \hat n}_0=-\bf{\hat{x}} &  {\bf \hat n}_4=\left({h{\bf\hat y}-\frac{p}{2}{\bf\hat z}}\right)/\sqrt{h^2+(p/2)^2}\\
{\bf \hat n}_1=\bf{\hat{x}} & {\bf \hat n}_5=\left({-h{\bf\hat y}-\frac{p}{2}{\bf\hat z}}\right)/\sqrt{h^2+(p/2)^2}\\
{\bf \hat n}_2=\bf{\hat{y}} & {\bf \hat n}_6=\left({h{\bf\hat x}-\frac{p}{2}{\bf\hat z}}\right)/\sqrt{h^2+(p/2)^2}\\
{\bf \hat n}_3=-\bf{\hat{y}} & {\bf \hat n}_7=\left({-h{\bf\hat x}-\frac{p}{2}{\bf\hat z}}\right)/\sqrt{h^2+(p/2)^2}.
\end{array}
\end{equation}

For a particular angle under consideration, many runs are done, each corresponding to a different position, $\bf{r}_0$, in the $z=0$ plane. For the square geometry presented here, a grid of starting positions is defined. The results of each run are averaged to determine the total response. 

In each case, the unit ${{\bf \hat k}}$ associated with the angle is used to define a direction. The polarization state of the vector is defined with respect to the $y$-axis. ${\bf{\hat e}}_1={{ \bf \hat y}}$. The orthogonal linear polarization is then 
\begin{align}
{{\bf \hat e}}_2={\bf \hat {y}}\times{{\bf \hat k}}.
 \end{align}
 In this way, Stokes $Q$ is defined as
 \begin{align}
 Q\equiv(\bf{E}\cdot{\bf{\hat e}}_1)^2-(\bf{E}\cdot{\bf{\hat e}}_2)^2,
 \end{align}
where $\bf{E}$ is the electric field ,which is orthogonal to ${\bf{\hat k}}$.

The distance to each of the 8 surfaces along this direction is determined, and the ``next surface'' is defined as that having the shortest distance, $d_{min}$, from the current position of the ray. Care has to be taken to define the finite extent of the pyramid's walls. These are not infinite planes, but rather planes that end when they intersect the other planes of the pyramid. The ray is propagated to this surface, 
\begin{align}
{\bf{r}}=d_{min}{\bf{\hat k}}+\bf{r}_0,
\end{align}
where $\bf{r}$ is the new position.

If the new position is on one of the boundary walls, the position of the vector is re-mapped to the opposite wall. That is, for surfaces 0 and 1, the $x$-coordinate is reflected about the origin. For surfaces 2 and 3, the $y$-coordinate is reflected about the origin.  

If the new position is on one of the pyramid walls, the interaction involves both determining the new direction of $\hat{\bf{k}}$ and modifying the polarized amplitudes according to the Fresnel equations. 

To handle the differing polarization responses, the amplitudes of the field along the $\hat{\bf{e}}_1$ and $\hat{\bf{e}}_2$ vectors are projected onto a basis parallel and perpendicular to the plane of incidence (and perpendicular to $\hat{\bf{k}}$.) The unit vectors parallel and perpendicular to the plane of incidence are
\begin{align}
{\bf{\hat e}}_1^\prime={\bf{\hat k}}\times\hat{\bf{n}}_i\\
{\bf{\hat e}}_2^\prime={\bf{\hat e}}_1^\prime\times{\bf{\hat k}}.
\end{align}
where $\hat{\bf{n}}_i$ is the unit normal vector to the surface. The angle, $\phi$, between this vector and the ${\bf{\hat e}}_1$ is given by
\begin{align}
\phi=\cos^{-1}({\bf{\hat e}}_1\cdot{\bf{\hat e}}_1^\prime).
\end{align}
The sign of this angle is determined by the sign of $\alpha={\bf{\hat k}}\cdot{\bf{\hat e}}_1\times{\bf{\hat e}}_1^\prime$. The angle $\phi$ is defined to be positive as measured counterclockwise (as viewed along $\bf{{\hat e}}_1$) from ${\bf{\hat e}}_1^\prime$, such that if $\alpha>0$, $\phi<0$.
The Stokes parameters in the new basis can then be found.

\begin{align}
\bf{S}^\prime &=R(\phi)\bf{S}\\
\left(\begin{array}{c}
I^\prime\\
Q^\prime\\
U^\prime\\
V^\prime\\ \end{array}\right)&=
\left(\begin{array}{cccc}
1 &0&0&0\\
0&\cos{2\phi}&\sin{2\phi}& 0\\
0&-\sin{2\phi}&\cos{2\phi}& 0\\
0&0&0&1\\
\end{array}\right)
\left(\begin{array}{c}
I\\
Q\\
U\\
V\\
\end{array}\right)
\label{eq:mueller}
\end{align}
The $Q$ direction is now defined by $Q\equiv(\bf{E}\cdot\hat{\bf{e}}_1^\prime)^2-(\bf{E}\cdot\hat{\bf{e}}_2^\prime)^2$. 

The next step is to incorporate the physical interaction between the incident radiation and the surface of the absorber structure. The Fresnel amplitude coefficients for reflections associated with the polarizations parallel and perpendicular to the surface are
\begin{align}
\Gamma_{\perp}&=\frac{\cos{\theta}-\sqrt{\epsilon_r}\cos{\theta_t}}{\cos{\theta}+\sqrt{\epsilon_r}\cos{\theta_t}}\\
\Gamma_{\parallel}&=\frac{\cos{\theta_t}-\sqrt{\epsilon_r}\cos{\theta}}{\cos{\theta_t}+\sqrt{\epsilon_r}\cos{\theta}}.
\end{align}
Here, $\epsilon_r$ is the complex relative permittivity and $\theta_t$ is the transmitted angle relative to normal incidence, which can be determined by
\begin{align}
\cos{\theta_t}=\sqrt{1-\frac{1}{\epsilon_r}\sin^2{\theta}}.
\end{align}
 The incidence angle is defined by $\theta=\cos^{-1}({\bf{\hat k}}\cdot{\bf{\hat n}}_i).$  

The polarization-dependent power reflection can be computed by multiplying the Stokes vector by the following matrix.
\begin{align}
\Gamma^2(\theta)=
\left(\begin{array}{cccc}
\Gamma_\Sigma^2  &\Gamma_\Delta^2&0&0\\
\Gamma_\Delta^2&\Gamma_\Sigma^2&0& 0\\
0&0&-\Gamma_\Delta^2& 0\\
0&0&0&-\Gamma_\Delta^2\\
\end{array}\right)
\end{align}
Here, we have defined
\begin{align}
\Gamma_\Sigma^2&\equiv\frac{1}{2}\left(\Gamma_{\perp}^2+\Gamma_{\parallel}^2\right) \\
\Gamma_\Delta^2&\equiv\frac{1}{2}\left(\Gamma_{\perp}^2-\Gamma_{\parallel}^2\right),
\end{align}
where $\Sigma$ and $\Delta$ specify the sum and difference in power reflection, respectively.
The negative signs for the last two diagonal elements are capturing the parity flip of $U$ and $V$ associated with the reflection.

The outgoing direction is determined by the law of reflection. This is done by first rotating ${\bf{\hat k}}$ around the normal to the surface through an angle of $\pi$ and then flipping the sign of ${\bf{\hat k}}$ so that the outgoing vector is directed away from the surface. The rotation is accomplished by utilizing the Euler-Rodrigues method.

\begin{figure} [p]
   \begin{center}
   \begin{tabular}{c} 
   \includegraphics[width=2.5in,angle=0]{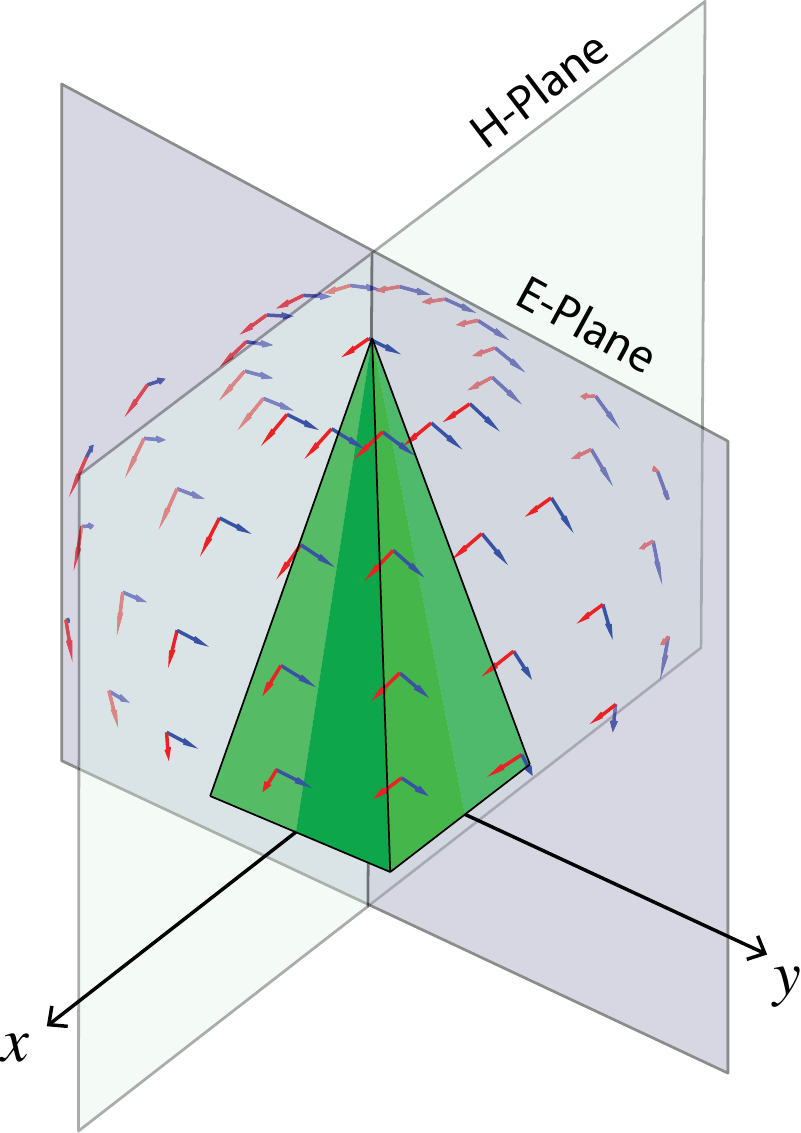}
   \end{tabular}
   \end{center}
   \caption[template] 
   { \label{fig:crosspol} 
   The coordinate system for the polarization direction is shown. Blue arrows represent the direction defined as parallel to the incident polarization (${\bf{\hat e}}_{\mathit{ref}}$). Red arrows indicate the perpendicular direction (${\bf{\hat e}}_{\mathit{cross}}$). The E- and H-Planes are shown. These correspond to the incident planes in Figure~\ref{fig:gmod1}.} 
   \end{figure}

This process is repeated for each ray until the ray returns to a negative $z$-coordinate. At this point, it is necessary to project the resulting polarization onto a basis common to all of the exiting rays for a given angle. To do this, we follow the formalism of Ludwig's third definition \cite{Ludwig73}, choosing the $y$-direction as our nominal ``co-polarization'' reference. We define orthogonal unit vectors as
\begin{align*}
&{\bf{\hat e}}_{\mathit{ref}}=-(1-\cos{\theta})\sin{\phi}{\bf{\hat x}}+[1-\sin^2{\phi}(1-\cos{\theta})]{\bf{\hat y}}\\
&-\sin{\theta}\sin{\phi}{\bf{\hat z}}\\
&{\bf{\hat e}}_{\mathit{cross}}=[1-\cos^2{\phi}(1-\cos{\theta})]{\bf{\hat x}}-(1-\cos{\theta})\sin{\phi}\cos{\phi}{\bf{\hat y}}\\
&-\sin{\theta}\cos{\phi}{\bf{\hat z}}.
\end{align*}
As before, the angle between the current definition of ${\bf{\hat e}}_1$  and ${\bf{\hat e}}_{\mathit{ref}}$ is determined by $\cos^{-1}({\bf{\hat e}}_1\cdot{\bf{\hat e}}_{\mathit{ref}})$, with the sign begin determined by ${\bf{\hat k}}\cdot{\bf{\hat e}}_1\times{\bf{\hat e}}_{\mathit{ref}}$. The Stokes vector can be rotated into this basis using Equation~\ref{eq:mueller}. The final basis set is then defined as ${\bf{\hat e}}_1={\bf{\hat e}}_{\mathit{ref}}$, ${\bf{\hat e}}_2={\bf{\hat e}}_{\mathit{cross}}$. Figure~\ref{fig:crosspol} shows this reference frame.
   
For each incidence angle, many such raytraces are performed and the fractional
Stokes parameters (relative to unity input power) are averaged to determine the total power in each polarization that are reflected from the device. The final Stokes vector for a given wave vector, $\bf{S}(\hat{\bf{k}})$ is computed by averaging over the set of rays,
\begin{align}
\langle{\bf{S}(\hat{\bf{k}})}\rangle=\frac{1}{N}\sum_{j=1}^{N}\left({R(\phi_j)\prod_{i=1}^{M} \Gamma^2(\theta_i)R(\phi_i)}\right)\bf{S}(\hat{\bf{k}}).
\end{align}

%


\end{document}